\documentclass[letterpaper,11pt]{article}

\usepackage[T1]{fontenc}
\usepackage[utf8]{inputenc}
\usepackage{authblk}
\usepackage{amsthm}
\usepackage{amsmath}
\usepackage{amsfonts}
\usepackage{verbatim}
\usepackage{tikz}
\usepackage{multirow}
\usepackage{algorithm2e}
\usepackage{graphicx}
\usepackage{caption}
\usepackage{subcaption}
\usepackage{amssymb}

\newtheorem{lemma}{Lemma}
\newtheorem{theorem}{Theorem}
\newtheorem{definition}{Definition}

\newcommand{\simf}{\stackrel{f}{\sim}}
\newcommand{\simp}{\stackrel{p}{\sim}}
\newcommand{\s}{\mathcal{S}}
\newcommand{\T}{\mathcal{T}}

\newcommand{\eqsi}{\s^i /\mathord{\sim}}

\newcommand{\eqsfi}{\s^i /\mathord{\simf}}

\newcommand{\eqspi}{\s^i /\mathord{\simp}}
\newcommand{\eqspii}{\mathcal{P}^{i} /\mathord{\simp}}

\title{ \vspace*{-0.5in} Timestamps for Partial Replication\thanks{This research is supported in part by National Science Foundation award 1409416, and Toyota InfoTechnology Center. Any opinions, findings, and conclusions or recommendations expressed here are those of the authors and do not necessarily reflect the views of the funding agencies or the U.S. government.}}
\author[1]{Zhuolun Xiang\thanks{xiangzl@illinois.edu}}
\author[2]{Nitin H. Vaidya\thanks{nhv@illinois.edu}}
\affil[1]{Computer Science Department}
\affil[2]{Department of Electrical and Computer Engineering\protect\linebreak University of Illinois at Urbana-Champaign}
\date{November 15, 2016 (revised December 26, 2016)\thanks{The revision mainly adds
Section 5.7, expands Section 5.6 of the previous version.}}

\begin{document}
  \maketitle

\begin{comment}
	+gentle rain
	compare to Meldal1991ExploitingLI
	approximation, fixed size timestamp
	
\end{comment}  

\begin{abstract}
	Maintaining causal consistency in distributed shared memory systems using vector timestamps has received a lot of attention from both theoretical and practical prospective. However, most of the previous literature focuses on full replication where each data is stored in all replicas, which may not be scalable due to the increasing amount of data.
	In this report, we investigate how to achieve causal consistency in partial replicated systems, where each replica may store different set of data. We propose an algorithm that tracks causal dependencies via vector timestamp in client-server model for partial replication. The cost of our algorithm in terms of timestamps size varies as a function of the manner in which the replicas share data, and the set of replicas accessed by each client.
	We also establish a connection between our algorithm with the previous work on full replication.
	%To show the optimality of our algorithm, we develop a lower bound for timestamp size in partial replication. We calculate the bound explicitly, and show our algorithm matches the lower bound.
\end{abstract}

\section{Introduction}

Geo-replicated system provides fault tolerance and low latency when processing large amount of data. Data replication is the most commonly used method to guarantee such properties. Multiple copies of the data are stored in different replicas located in various regions, providing efficient communications with clients who want to access the data. However, the replication of data introduces another problem, the consistency of the data replicas. Various consistency models have been proposed and investigated in previous literature, including linearizability, sequential consistency, causal consistency and eventual consistency. Linearizability is the strongest form of consistency which asks the operations must be linearizable, while eventual consistency is the weakest one, requiring the data to be eventually consistent. Among these consistency models, causal consistency is receiving increasing attention recently.

Causal consistency provides a causally consistent view of the clients, by defining and enforcing causal dependencies between events. It is well-known that causal consistency is the strongest form of consistency model which provides low latency \cite{Lloyd2011DontSF}. Therefore, an increasing number of systems provide causal consistency, such as COPS \cite{Lloyd2011DontSF}, Orbe \cite{Du2013OrbeSC}, SwiftCloud \cite{Zawirski2014SwiftCloudFG} and GentleRain \cite{Du2014GentleRainCA}. The above systems provide causal consistency with low latency, but they may not be scalable due to the fact they assume full replication. When the amount of data increases, each replica has to maintain a copy of all the data, which would be unrealistic due to the explosion of big data nowadays. Facing this challenge, partial replication is a promising way to solve the problem.

Compared to the large number of protocols assuming Complete Replication and Propagation (CRP), partial replication has received less attention. Several researchers have addressed challenges in achieving causal consistency in partial replication, mainly because of the large amount of metadata the system needs to keep track of in order to characterize the accurate dependencies \cite{Bailis2012ThePD,Lloyd2011DontSF,Hlary2006AboutTE, crain2015designing}. 
Tracking causal dependencies with minimum amount of metadata is an interesting problem from both theoretical and practical prospective.

The contributions of this paper can be summarized as follows:

\begin{enumerate}
	\item We propose an algorithm that utilizes the notion of share graph to capture the causal dependencies in partial replication systems via vector timestamps.
	\item Our algorithm characterizes the effect of clients' communication patterns with the replicas on timestamps sizes. The cost of keeping causal consistency varies with clients accessing different subset of replicas.
	\item We establish a connection between the solutions for partial replication and full replication. Our algorithm degenerates gracefully from applying to partial replication to full replication.
\end{enumerate}

\section{Related Work}
\subsection*{Message passing}
Lamport timestamp was first proposed to order events according to \textsl{happened-before} relationship in distributed systems \cite{lamport1978time}. Fidge and Mattern proposed a partial ordering characterization using vector clock \cite{Mattern1988VirtualTA,fidge1987timestamps}.

Rodrigues and Ver{\'i}ssimo reduced the timestamps cost in message passing systems using the information about the communication topology \cite{rodrigues1995causal}. 

Meldal et al. also reduced the timestamps sizes in message passing model, by exploiting the topology of the communication graph, and only ordering events arriving at the same process \cite{Meldal1991ExploitingLI}. This shares some similarities with maintaining causal consistency in shared memory, because all we care about is maintaining update orders at each replica.

\subsection*{Full replication}
Lazy Replication \cite{Ladin1992ProvidingHA} is a classic framework for providing causal consistency in distributed shared memory, where all clients and replicas both maintain a vector clock of size equal to the number of replicas. Clients can issue updates and queries to any replica, and replicas exchange gossip messages to keep their data up-to-date. 

COPS implements a causally consistent key-value store system \cite{Lloyd2011DontSF}, that computes a list of dependencies whenever a update occurs, and the operation is not performed until updates in the dependencies are applied.
The size of dependency list is reduced using the transitivity rule of happened-before relationship.

SwiftCloud provides efficient writes and reads using an occasionally stale but causally-consistent client-side local cache scheme \cite{Zawirski2014SwiftCloudFG}. Similar to Lazy Replication, the size of the metadata in SwiftCloud is proportional to the number of data centers used to store the data.

Orbe is a causally consistent distributed key-value storage system \cite{Du2013OrbeSC}. Two protocols are used to provide causal consistency: the DM protocol which uses two-dimensional matrices to track the dependencies, and the DM-Clock protocol which provides read-only transactions via loosely synchronized physical clocks.

\subsection*{Partial replication}

Milani systematically studied achieving causal consistency in distributed systems in her PhD thesis \cite{Xphdthesis}.
Milani et al. proposed an optimality criterion for a protocol to achieve causal consistency under relation $\rightarrow_{co}$ in full replication. They also presented a corresponding optimal protocol that can apply the update as soon as possible \cite{Baldoni2006OptimalPP}. 
H{\'e}lary and Milani proposed the idea of share graph in partial replication, and they defined the set of \textsl{$x$-relevant} processes, which are the processes storing $x$ or belonging to a \textsl{minimal $x$-hoop} \cite{Hlary2006AboutTE}. By revealing that \textsl{$x$-relevant} processes must carry dependencies of $x$ in their metadata, H{\'e}lary and Milani identified the difficulty of efficient implementation under causal consistency.
They also studied several weakened versions of consistency models between causal consistency and PRAM \cite{raynal1995causal}, such as \textsl{Lazy causal consistency} and \textsl{Lazy semi-causal consistency}. However, they discovered that only PRAM allows efficient partial replication implementations among the above consistency models .

Crain and Shapiro addressed challenges on maintaining causal consistency in partial replicated system. They proposed a protocol that costs less than full replication, but enforces some false dependencies \cite{crain2015designing}. The key insight of their protocol is to let the sender check the dependencies instead of the receiver when propagating updates among data centers. This approach avoids unnecessary dependency checks caused by partial replication, with the cost that the observed data may be more stale.

Shen et al. \cite{Shen2015CausalCF} studied achieving causal consistency under relation $\rightarrow_{co}$ proposed by Milani et al. \cite{Baldoni2006OptimalPP}.
In their model, clients communicate with a single site, and fetch data from other predesignated site if the data is not locally stored. The sites are connected by FIFO channels.  They proposed two algorithms, \textsl{Full-Track} and \textsl{Opt-Track}. The first one characterizes accurate dependencies, with message size cost $O(n^2)$ where $n$ is the number of sites. 
The second one adapts \textsl{KS algorithm} \cite{Kshemkalyani1998NecessaryAS} to this setting, and reduces amortized message size to $O(n)$ and worst case message size is $O(n^2)$.

Kshemkalyani and Hsu's research on approximate causal consistency followed the system model in \cite{Shen2015CausalCF}, and sacrificed accuracy of causal consistencies to reduce the meta-data \cite{Kshemkalyani2015ApproximateCC}.

\section{System Model}\label{sec:model}
In this section, we introduce the system model formally. We consider a client-server model \cite{tanenbaum2007distributed}, where each client can communicate with a fixed subset of replicas. The notions of share graph \cite{Hlary2006AboutTE} and causal consistency \cite{attiya2004distributed} are defined.

\subsection{Client-server model}
We consider the traditional client-server model \cite{tanenbaum2007distributed}. We will use the term replica and server interchangeably in the remaining context.
The system consists $n$ replicas $r_1,\cdots,r_n$ and $m$ clients $c_1,\cdots, c_m$. 
We will use the terms replica $i$ and replica $r_i$, client $i$ and client $c_i$ interchangeably.
The communication graph of replicas is a complete graph. Communication channels are reliable but not FIFO. Each client $i$ can communicate with a fixed subset of replicas $R_i$. Each client can perform update/query (write/read) operations on variables stored by the replicas they communicate with.

\subsection{Partial replication}
\begin{figure}[h]
	\centering
	\includegraphics[width=0.4\textwidth]{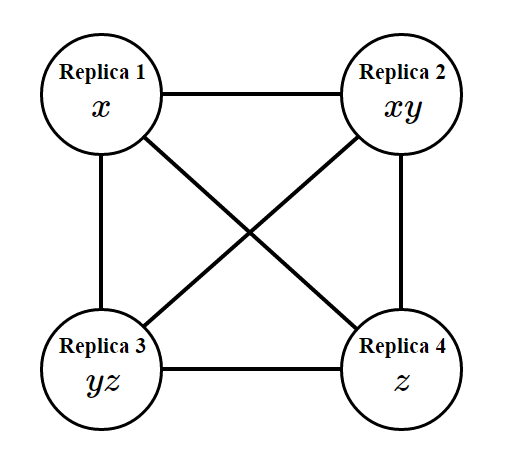}
	\caption{Partial replication}
	\label{fig:pr}
\end{figure}

Each replica $i$ stores (or "owns") a set of data $X_i$. In contrast to previous full replication literature where $X_i$s are identical at all replica $i$, we consider partial replication where $X_i$ can be different at different replicas. As a simple example, in Figure \ref{fig:pr}, replica $1$ owns data $x$, replica $2$ owns data $x,y$, replica $3$ owns data $y,z$, and replica $4$ owns data $z$.

\subsection{Share graph}
\begin{figure}[h]
	\centering
	\includegraphics[width=0.4\textwidth]{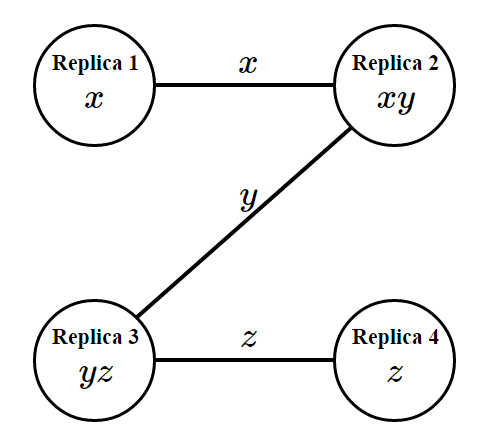}
	\caption{Share graph}
	\label{fig:sg}
\end{figure}

The notion of share graph was proposed by H{\'e}lary and Milani \cite{Hlary2006AboutTE} in order to show the difficulty of tracking causal dependencies in partial replicated systems. 
We will use same notion of share graph as defined below. A simple example of share graph corresponding to Figure \ref{fig:pr} is given in Figure \ref{fig:sg}.
\begin{definition}[Share Graph]
	A share graph is undirected graph $G=(V,E)$, where $V$ is the set of all replicas, and an  edge $e_{i,j}\in E$ if and only if $X_i\cap X_j\neq \emptyset$.
\end{definition}

\subsection{Data consistency}\label{sec:data_consistency}
We are interested in keeping data causally consistent in the partial replication system.
Causality is first defined by Lamport as \textsl{happened-before} relationship between two operations \cite{lamport1978time}. Let $o_1,o_2$ be two operations, we denote $o_1\rightarrow o_2$ as $o_1$ happens before $o_2$. We say $o_1$ happens before $o_2$, or $o_2$ depends on $o_1$, if and only if one the following three conditions is true:
\begin{enumerate}
	\item Program order: Both $o_1$ and $o_2$ are performed by a single process, and $o_1$ happens before $o_2$.
	\item Read from: $o_1$ is a write operation, and $o_2$ is a read operation that reads the value written by $o_1$.
	\item Transitivity: There exist another operation $o_3$ that $o_1\rightarrow o_3$ and $o_3\rightarrow o_2$.
\end{enumerate}

\begin{definition}[Causal consistency]
	A system is causally consistent if each operation is applied only after the causal dependencies of that operation are all reflected in the system; also, when the dependencies of some operation are all reflected in the system, the operation will eventually be applied.
\end{definition}

\section{Difficulties in partial replication}
\begin{figure}[h]
	\centering
	\includegraphics[width=0.8\textwidth]{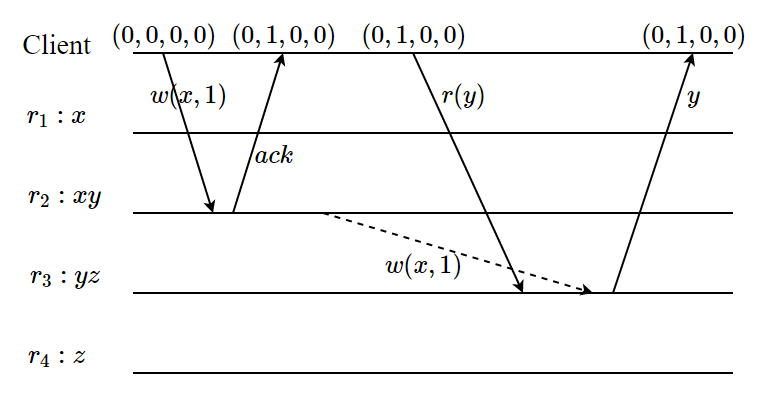}
	\caption{False-dependency using previous method}
	\label{fig:dif}
\end{figure}

Timestamping methods designed for full replication will introduce lots of false-dependencies when maintaining causal consistency of the data in partial replication systems \cite{crain2015designing}. See Figure \ref{fig:dif} as an example, where the client communicates with replica $r_1,r_2,r_3$ and $r_4$ depicted in Figure \ref{fig:pr}. A similar example is shown in \cite{crain2015designing}.

According to Lazy Replication or similar method for full replication, both client and replicas will keep a vector clock of size equal to the number of replicas. All clocks are $0$ initially. First the client performs a write operation on $x$ to replica $2$. The timestamps returned by the replica is $(0,1,0,0)$, indicating this write operation is in the causal past. Then the client performs a read operation on $y$ to replica $3$, before the gossip message from replica $2$ arrives at replica $3$. Since replica $3$ cannot tell whether the read operation causally depends on the previous operation at replica $2$, it will wait until the gossip message arrives. However, this scheme introduces false-dependencies, which increases the latency of system dramatically. To accurately track the causal dependencies, larger timestamps are needed.

\section{Algorithm}\label{sec:algo}
In this section, we present our algorithm for maintaining causal consistency in partial replication. Our scheme is based on Lazy Replication for the client-server model \cite{Ladin1992ProvidingHA}. We focus on the timestamps sizes of replicas and clients, and omit the garbage collection part in the algorithm for brevity.

\subsection{Augmented Share Graph}
Before presenting the algorithm, we characterize the augmented share graph to calculate timestamp sizes used in our algorithm. The structure of our vector timestamp is inspired by previous work of H{\'e}lary and Milani on partial replication \cite{Hlary2006AboutTE}, which is also mentioned in the section of Related Work. Essentially, they proposed the idea of $x$-relevant processes, which characterizes the set of processes that have to store the dependencies of operations on $x$. While no specific timestamp structure is provided in their work, we describe an algorithm that uses edge-based vector timestamps. The proposed timestamps are intuitive and easy to compute. In addition, our contributions include the notion of {\em augmented} share graph, and as discussed later, identifying a connection between partial replication and full replication.

Suppose client $c_i$ communicates with replicas in the set $R_i$ where $i=1,\cdots, m$. The augmented share graph is defined as follows. 

\begin{definition}[Augmented Share Graph]
	An augmented share graph $G'=(V',E')$ of a share graph $G=(V,E)$ is an undirected graph, where $V'=V$, and $E'=E\cup \{e_{s,t}~|~\exists c_i, ~r_s\in R_i,r_t\in R_i\}$
\end{definition}

Essentially, the augmented share graph is created based on clients' communication patterns and the share graph.
For every client $c_i$, let $r_s,r_t$ be any two replicas in $R_i$, we add edge $e_{s,t}$ in the share graph. After the above procedure we obtain the augmented share graph $G'$. 

\begin{figure}
	\centering
	\begin{subfigure}[b]{0.52\textwidth}
		\includegraphics[width=\textwidth]{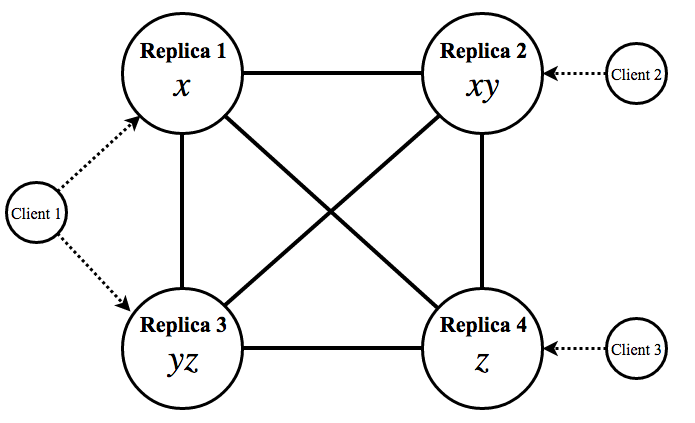}
		\caption{Clients access replicas}
		\label{fig:pr2}
	\end{subfigure}
	~ %add desired spacing between images, e. g. ~, \quad, \qquad, \hfill etc. 
	\begin{subfigure}[b]{0.35\textwidth}
		\includegraphics[width=\textwidth]{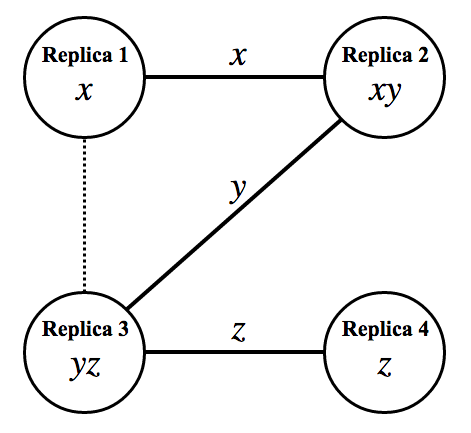}
		\caption{Augmented share graph}
		\label{fig:asg}
	\end{subfigure}
	\caption{Example of augmented share graph}\label{fig:example}
\end{figure}

Figure \ref{fig:example} contains a simple example of augmented share graph. Figure \ref{fig:pr2} is the same example as Figure \ref{fig:pr} with clients communicating the replicas, and Figure \ref{fig:asg} is the corresponding augmented share graph. Here, $R_1=\{r_1, r_3\}$, $R_2=\{r_2\}$ and $R_3=\{r_4\}$.

Consider the augmented share graph $G'$ and replica $i$ in the graph. Let $L'_i$ be the set of edges in all simple loops that pass through $i$ in $G'$. Let $L_i=L'_i\cap E$.
Let $N'_i$ denote the set of edges from replica $r_i$ to its neighbors in $G'$. Let $N_i=N'_i\cap E$.
Define $E'_i=L_i\cup N_i$. 
As we will see soon, our timestamping method keeps counters for each direction of the edges in set $E'_i$ of replica $i$.
For convenience, we will refer each undirected edge $e_{s,t}$ as two directed edges $e_{st}$ or $e_{ts}$, and let $E_i$ denote the set of directed edges corresponding to the undirected edges in $E'_i$. That is, 
$$E_i=\{e_{st}~|~\forall e_{s,t}\in E'_i\} ~\bigcup ~ \{e_{ts}~|~\forall e_{t,s}\in E'_i\}$$
Define $C_i=\bigcup_{j\in R_i} E_j$, which is the set of directed edges client $i$ needs to keep counters for. 

As an example, we characterize the sets $E_i$ and $C_i$ in Figure \ref{fig:example}. We denote the directed edge from $r_i$ to $r_j$ as $e_{ij}$. Then $E_1=E_2=\{e_{12},e_{21},e_{23},e_{32} \}$, $E_3=\{e_{12},e_{21},e_{23},e_{32}, e_{34},e_{43} \}$, $E_4=\{e_{34},e_{43} \}$, $C_1=E_1\cup E_3= \{e_{12},e_{21},e_{23},e_{32}, e_{34},e_{43} \}$, $C_2=E_2=\{e_{12},e_{21},e_{23},e_{32} \}$, $C_3=E_4=\{e_{34},e_{43} \}$.

When we say $x\in e_{st}$ where $e_{st}$ is an edge between replica $s$ and $t$, it means $x$ is shared between replica $s$ and $t$. For example, $x\in e_{12}$ and $x\in e_{21}$ in Figure \ref{fig:asg}.

\subsection{Notations}
We define the timestamps and operations used in the algorithm as follows. The structure of our vector timestamp is inspired by the previous work of H{\'e}lary and Milani mentioned in the Related Work section \cite{Hlary2006AboutTE}. Their notion of $x$-relevant processes characterizes the set of processes that have to carry the dependency information of $x$ in their timestamps, which is similar to containing "simple loops" in our vector timestamps. One of the contributions of our scheme is how to enforce the dependencies when clients can communicate with multiple replicas.

\begin{enumerate}
	\item $c_i:$ timestamps of client $i$, recording the number of operations on shared variables on each edge seen by the client so far.
	Timestamp is a pair $(e_{xy}, k)$, where $e_{xy}\in C_i$ and $k$ is a sequence number.
	$c_i=\{(e_{xy}, k)~|~e_{xy}\in C_i \}$ is a set of timestamps $(e_{xy}, k)$,
	and contains $|C_i|$ timestamps, with each element corresponding to a directed edge in $C_i$. 
	
	For brevity, we will refer the sequence number $k$ corresponding to $e_{xy}$ simply as $c_i[e_{xy}]$.
	
	$rep_i$, $val_i$, $u_i.dep$, $q_i.dep$ and $r.dep$ defined below will follow the same notation as $c_i$ (i.e., in timestamp $(e_{xy},k)$).  
	
	Let $c_i|_{r_j}=\{(e, k)\in c_i ~|~ e\in E_{j} \}$ denote the components of $c_i$ in the intersection of client's timestamps with replica $i$'s timestamps (defined below).
	
	\item $rep_i:$ timestamps of replica $i$, recording the number of operations on shared variables on edges that is contained in the log. 
	$rep_i=\{(e_{xy},k)~|~ e_{xy}\in E_i \}$ is a set of timestamps $(e_{xy}, k)$, where $e_{xy}\in E_i$ and $k$ is a sequence number. $rep_i$ contains $|E_i|$ timestamps, with each element corresponding to a directed edge in $E_i$. 
	
	\item $val_i:$ timestamps of replica $i$, recording the number of operations already applied. $val_i$ have the same structure and size as $rep_i$.
	
	\item $u_i(x,v,dep):$ update message on replica $i$ for a write of value $v$ to $x$. $u_i.dep$ has the same structure and size as $rep_i$. 
	
	\item $q_i(x,dep):$ query message on replica $i$ for a read of $x$. $q_i.dep$ has the same structure and size as $rep_i$. 
	
	\item $r(id, x,v,ts,dep):$ record of update message stored in replica $i$'s log for an operation that writes $v$ to $x$. 
	$id$ is the identifier of the replica at which a client invoked this operation.
	$r.ts$ and $r.dep$ have the same structure as $rep_i$.
	
	\item $Done:$ the set of writes that have been applied.
\end{enumerate}

We will omit the subscript $i$ of the above notations when there is no ambiguity in the algorithm.

In the following algorithm, let $a,b$ be two timestamps, and $E^a,E^b$ be the corresponding set of edges that $a,b$ contain.
We define $c=merge(a,b)$ where $E^c=E^a$ and
\begin{equation*}
c(e)=\left\{
\begin{aligned}
&\max(a(e),b(e)),  &&e\in E^a\cap E^b\\
&a(e),  &&e\in E^a\backslash E^b\\
\end{aligned}
\right.
\end{equation*}
Note that $merge(a,b)$ is not necessarily equal to $merge (b,a)$. In particular, $merge(a,b) =merge(b,a)$ if and only if $E^a = E^b$.

\subsection{Client's algorithm}
Replica $i$ that a client $c_p$ may communication with may be any replica in set $R_p$.
\subsubsection*{Sending an update message}
When issuing an update on $x$ (i.e, write) to replica $i$, the client sends $u=(x,v, c|_{r_i})$ to replica $i$.

\subsubsection*{Receiving a reply of update message}
When receiving a reply of update with timestamp $t$, the client merges the received timestamps $t$ with its timestamps $c$, namely $c=merge(c,t)$.

\subsubsection*{Sending a query message}
When issuing a query on $x$ (i.e., read) to replica $i$, the client sends $q=(x, c|_{r_i})$ to replica $i$

\subsubsection*{Receiving a reply of query message}
When receiving a reply of query with timestamp $t$, the client merges the received timestamps $t$ with its timestamps $c$, namely $c=merge(c,t)$.

\subsection{Replica's algorithm}
\subsubsection*{Processing an update message $u=(x,v, dep)$}
When replica $i$ receives an update message $u=(x,v, dep)$, it does the following:

\begin{algorithm}[H]
	// Advance the replica's timestamp: \\
	$rep[e_{ij}]=rep[e_{ij}]+1$, $\forall e_{ij}\in E_i$ such that $x\in e_{ij}$\\
	// Compute the timestamp for the update: \\
	
	$ts=u.dep$, $ts[e_{ij}] = rep[e_{ij}]$, $\forall e_{ij}\in E_i$ such that $x\in e_{ij}$\\
	// Construct the update record $r$ and add to log: \\
	$r=(i,~x, ~u.v, ~ts, ~u.dep)$\\
	add $r$ to the log\\
	\If{$u.dep[e_{ki}]\leq val[e_{ki}]$, $\forall e_{ki}\in E_i$} {
		// Apply the write: \\
		$x=u.v$\\
		
		$val=merge(val,r.ts)$ \\
		
		$Done=Done\bigcup \{r\}$\\
	}
	return $r.ts$ in a reply message
\end{algorithm}

\subsubsection*{Processing a query message $q=(x, dep)$}
When replica $i$ receives a query message $q=(x, dep)$, it does the following:

\begin{algorithm}[H]
	\textbf{wait}
	\textbf{until} $q.dep[e_{ki}]\leq val[e_{ki}]$, $\forall e_{ki}\in E_i$\\
	return the value of $x$ and $val$
\end{algorithm}

\subsubsection*{Sending a gossip message}
Replica $j$ only need to send gossip messages to its neighbors in the share graph $G$. 
The gossip message $m$ sent to a neighbor replica, say replica $i$, contains sender $j$'s timestamp $m.ts=rep$, and the update records from sender $j$'s log that apply on data shared by $i$ and $j$, namely $m.log=\{r\in log ~|~ r.x\in e_{ji} \}$.
%And each record needs to be sent only once since we assume no failures or message losses.

\subsubsection*{Processing a gossip message}
When replica $i$ receives a gossip message $m=(ts, log)$ from replica $j$, it does the following:

\begin{algorithm}[H]
	// Add the log to the replica's log: \\
	$log=log\cup m.log$\\
	// Merge the timestamps: \\
	$rep=merge(rep, m.ts)$\\
	// Select the update records that are ready to apply: \\
	$comp=\{r\in log ~|~ r.dep[e_{ki}]\leq rep[e_{ki}], \forall e_{ki}\in E_{r.id}\}$\\
	// Compute the value:\\
	\While{$comp$ not empty}{
		select $r$ from $comp$ such that $\not\exists r'\in comp$, s.t. 
		$\forall e_{ki}\in E_{r.id}$, $r'.ts[e_{ki}]\leq r.dep[e_{ki}]$
		
		$comp=comp-\{r\}$\\
		\If{$r \notin Done$} {
			// Apply the write:\\
			$x=r.v$\\
			$Done=Done\cup \{r\}$\\
		}
		$val=merge(val, r.ts)$
		
	}
\end{algorithm}

\subsection{Correctness}
We provide intuitive explanations why the algorithm is correct in the sense that it maintains causal consistency. Rigorous proofs will be presented in a future revision of this report.

\subsubsection*{Safety}
To show causal consistency is achieved by the above algorithm, we need to show that every update or record in the log is applied only after the updates that they causally depend on are reflected in the replica. We will intuitively describe how the algorithm tracks dependencies using the timestamps.

Firstly consider the dependencies introduced by clients. When a client issues update operation on a variable shared in some edges, the timestamps returned by the operation will update the corresponding components of client's timestamps to the latest value, and other components remain unchanged. 
Client's timestamp will merge with the current $val$ of the replica after a query operation, which adds the dependencies into client's causal history.
Therefore intuitively, the client's timestamps will be monotonically increasing and also capture the dependencies of all its update operations.

Secondly consider the dependencies created by gossip messages. This type of dependency is also addressed in \cite{Hlary2006AboutTE}, where the authors show the difficulty of efficiently implementing causal consistent shared memory in partial replication.

\begin{figure}[h]
	\centering
	\includegraphics[width=0.8\textwidth]{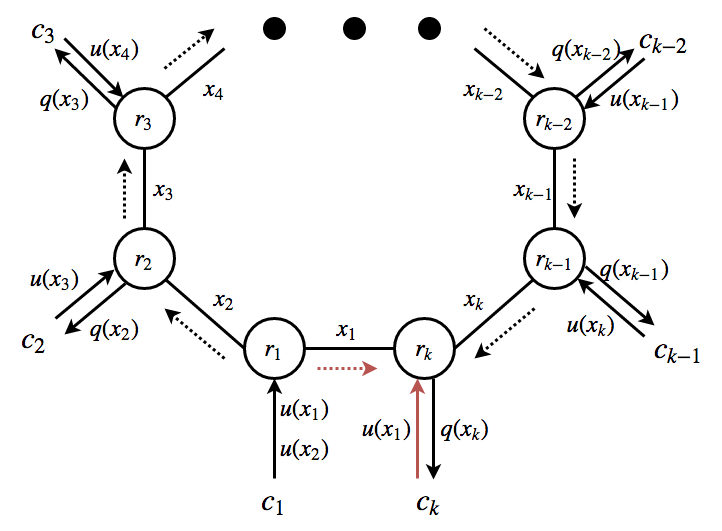}
	\caption{Loop in the share graph}
	\label{fig:loop}
\end{figure}

Refer to Figure \ref{fig:loop} as an example. Replica $r_1,r_2,\cdots, r_k$ form a simple loop in the share graph. Consider a procedure as follows. Client $c_1$ first issues an update $u(x_1)$ on replica $r_1$, then an update $u(x_2)$ on replica $r_1$. After the gossip message of update $u(x_2)$ arrives at replica $r_2$, client $c_2$ issues a query $q(x_2)$, and then an update $u(x_3)$. Similarly, after the gossip message of $u(x_3)$ arrives at replica $r_3$, client $c_3$ performs a query and an update. Repeating the process, when client $c_k$ issues update $u(x_1)$ (red arrow), it is easy to see that this update causally depends on the update $u(x_1)$ issued by client $c_1$. Therefore, it has to wait until the gossip message (dotted red arrow) arrives at $r_k$ and is applied.

In our algorithm, the replicas in this simple loop have timestamps containing components corresponding to all the edges in the simple loop. 
Whenever a client issues a query, the timestamps returned by the query will contain the dependencies.
Hence the dependency set in the above example is preserved along the propagation of the gossip messages.

\subsubsection*{Liveness}
We also need to show the liveness property, namely there is no deadlock in our algorithm.
For the client's operation, update is non-blocking, we only need to guarantee that query eventually returns.
Intuitively, every operation that a query depends on will eventually enter the log, and be applied. Therefore, the query eventually returns.

\subsection{Discussion}

\subsubsection*{Clients' communication pattern}
Clients' communications with replicas affect our timestamps cost significantly. 
As the construction of augmented share graph suggests, when a client communicates with two replicas who share no common data, it introduces extra dependencies between the data of these two replicas, which is equivalent to adding an edge in the share graph.
For example in Figure \ref{fig:pr2}, client $1$ first writes $x$ in replica $r_1$, and then writes $y$ in replica $r_2$. By the definition of causal relationship, the first write happens causally before the second one. 

The above dependencies introduced by clients' communications with multiple replicas increase the timestamps size the algorithm requires for both clients and replicas. We have the following observations:
\begin{enumerate}
	\item When each client only communicates with a single replica, then the timestamps size for both replica and client depends only on the original share graph. 
	
	\item When clients communicate with multiple replicas, the timestamps size for replicas depends on augmented share graph, which may have extra loops compared with share graph. Therefore, the timestamps size is very likely to increase. Nevertheless, the clients need to keep timestamps that are union of those replicas' timestamps they access.
	
	\item When clients communicate with all replicas, the timestamps sizes for all replicas and clients are the same, which equal to the number of all directed edges in the share graph. This is because by clients accessing all replicas, all edges in the share graph are in the loops in the augmented share graph.
\end{enumerate}
Hence, the timestamps cost increases when client communicates with more replicas. 

\subsubsection*{Comparison with full replication}
Intuitively, our algorithm eliminates the false dependencies such as the example given in the previous section, with the cost of larger timestamp size. More specifically, we store a counter per edge in the share graph that are necessary, instead of previous counter per replica method. Our method can be further optimized if multiple neighboring edges in the share graph actually have the same share variables on them. As an example, when replica $1$ shares data $x,y$ with both replica $2$ and $3$, only one counter for both neighboring edges $e_{12}, e_{13}$ is needed, because corresponding component of these two edges in the timestamps are always identical. In the extreme case of full replication, all neighboring edges have the identical set of variable. Therefore, only one counter per replica is needed in this case. Then it is easy to see that our algorithm becomes similar to the previous work in full replication. In other words, our timestamping method bridges partial replication with full replication, and edge-based timestamps degenerate to replica-based timestamps gracefully.

\subsubsection*{Multigraph structure for share graph}
When we treat the data shared in one edge as one part, the dependencies tracked by our algorithm are accurate. However, in general if the client requires higher level of accuracies, the timestamp sizes required maybe larger. For example, the algorithm may need to keep a counter per variable on an edge in the share graph, in order to keep track of their dependencies accurately.

In the discussion so far we assumed that the share graph is a simple graph.
Our approach can be readily extended to the case when multiple (virtual) links may be included between pairs of
replicas, with the data shared by the replicas connected by multiple links being partitioned across different
links between the replicas. For instance, suppose that replicas $1$ and $2$ share data $x,y,z$ in common, and we include
two links in the share graph between replicas $1$ and $2$. Then, for instance, variable $x$ may be assigned
to one of these two links, and variables $y,z$ being assigned to the other link. When variable $x$ is updated at replica $1$,
the timestamp element corresponding to the first of these links is updated. It is easy to conceive
other generalizations of the share graph, which trade off timestamp overhead and false dependencies
caused by the use of those timestamps.

\subsubsection*{Garbage collection}
Every replica keeps two sets locally, namely log and Done. Similar to \cite{Ladin1992ProvidingHA}, we can garbage collect log and Done in order to reduce the storage cost of the algorithm. If we assume no failures or message losses, the garbage collection is simple. Every record $r$ in the log can be removed, once it is already applied locally and sent to other replicas who also share the data that $r$ is applied on. Every record $r$ in Done can be removed, once the replica has received record $r$ from all possible neighbors in the share graph. 

More involved garbage collection similar to \cite{Ladin1992ProvidingHA} will be provided in the future version where we also consider failures and message losses.

\subsection{Reducing timestamps size via redundancy}\label{sec:reduce_size}
The timestamps used in the algorithm from previous section is simple and effective, which store a counter for each edge in the union of simple loops and neighbors. We may further reduce the size of the timestamps, if there exists redundancy among variables stored in neighbor edges of one replica. We elaborate this idea by giving the following definitions.

For replica $r_i$ in the augmented share graph $G=(V,E)$, define the set of directed neighbor edges $N^e_i=\{e_{ik},e_{ki} ~|~e_{i,k} \in E \}$ and the set of neighbor nodes $N^n_i=\{k ~|~ e_{i,k} \in E\}\cup \{i\}$. 
Define the set of simple loop edges $L^e_i$ as the set of directed edges in simple loops that pass through $i$. Define the set of simple loop nodes $L^n_i$ as the set of nodes in simple loops that pass through $i$.
Let $E_i=N^e_i\cup L^e_i$, $V_i=N^n_i\cup L^n_i$.

For replica $j\in V_i$, define $E_i^j=E_i\cap N^e_j$, which is the set of $j$'s neighbor edges that are in $E_i$. For each $e_{jk}\in E_i^j$, let $X_{jk}$ denote the set of variables shared by replica $j$ and $k$. Let $X_j=\bigcup_{e_{jk}\in E_i^j} X_{jk}$. Then for each $X_{jk}$, we assign a $|X_j|$--dimensional vector $\alpha_{jk}$, with each element $\alpha_{jk}[x]$ corresponding to a variable $x$ in $X_j$. For $x\in X_j$, if $x\in X_{jk}$, $\alpha_{jk}[x]=1$, otherwise $\alpha_{jk}[x]=0$. Denote the set of all $\alpha_{jk}$s as $\mathcal{A}$. Let $\tau_j$ be the maximum number of linearly independent vectors in $\mathcal{A}$. Clearly $\tau_j\leq |X_j|$. Let $\{\alpha_1,\alpha_2,\cdots,\alpha_{\tau_j}\}$ be one possible set of maximum linearly independent vectors in $\mathcal{A}$, and $\mathcal{E}^j_i=\{e_{1},e_{2},\cdots,e_{\tau_j}\}$ be the corresponding edges. Let $\mathcal{E}_i=\bigcup_{j\in V_i} \mathcal{E}_i^j$. We will show that it is enough for replica $i$ to store counters for each edge in $\mathcal{E}_i$ instead of $E_i$.

By definition, 
for any edge $e\in E_i^j\backslash\mathcal{E}_i^j$ where $\backslash$ denotes the set difference, 
let $\alpha_e$ denote the corresponding vector of $e$, then we have $\alpha_e=\sum_{p=1}^{\tau_j}a_p\alpha_p$ for some scalars $a_1,\cdots, a_p$. Let $U$ be a vector whose elements count the number of updates issued on corresponding variables in $X_j$. Then $U^T \alpha_e$ calculates the number of updates issued on variables shared on edge $e$. Let $ts[e]$ denote the number of updates issued on edge $e$, we have
$$
	ts[e]=U^T \alpha_e=U^T\sum_{p=1}^{\tau_j}a_p\alpha_p=\sum_{p=1}^{\tau_j}a_p(U^T\alpha_p)=\sum_{p=1}^{\tau_j}a_pts[e_p]
$$

Hence $ts[e]$ is a linear combination of $ts[e_p]$ where $e_p\in \mathcal{E}_i^j$.
Therefore it is enough for each replica to store counters for each edge in $\mathcal{E}_i$ to record the number of updates for all edges in $E_i$.
That is, replica $r_i$ only needs to maintain a timestamp of size equal to $\mathcal{E}_i$ instead of $E_i$.
When the above redundancy among the edges in $E_i$ is high, or the size of $\mathcal{E}_i$ is much smaller than $E_i$, the saving on timestamp sizes could be significant.

\section{Summary}
We proposed an algorithm that utilizes the notion of share graph to capture causal dependencies in partial replication systems via vector timestamps.
The cost of keeping causal consistency in our algorithm varies with different clients' communication patterns with the replicas.
We identify a connection between the solutions for partial replication and full replication, and our algorithm degenerates gracefully from applying to partial replication to full replication.
%We also develop a lower bound for timestamp size in full replication and partial replication. We calculate the bound explicitly, and show our algorithm matches the lower bound.

%ACKNOWLEDGMENTS are optional
%	\section{Acknowledgments}
	
	\bibliographystyle{abbrv}
	\bibliography{ref}  % sigproc.bib is the name of the Bibliography in this case

% +++++++++++++++++++++++++++++++++++++++++++++++++++++++++++

\appendix

\end{document}